\documentclass[%
 reprint,
superscriptaddress,
 amsmath,amssymb,
 aps,
floatfix
]{revtex4-2}

\usepackage{graphicx}
\usepackage{dcolumn}
\usepackage{bm}
\usepackage{siunitx}
\usepackage{float}
\usepackage{multirow}
\usepackage{color}
\usepackage{enumitem}
\usepackage{mathtools}
\usepackage{hyperref}
\usepackage{optidef}
\usepackage[capitalize]{cleveref}
\graphicspath{{./}{./figures/}{./figures/paper/}}
\newcommand{\norm}[1]{\left\lVert#1\right\rVert}
\newcommand{\pder}[2][]{\frac{\partial#1}{\partial#2}}
\newcommand{\taur}{\tau_{\mathrm{R}}}
\def\<#1\>{\mathinner{\langle#1\rangle}}

\newcommand{\nocontentsline}[3]{}
\newcommand{\tocless}[2]{\vspace{0.1in}\bgroup\let\addcontentsline=\nocontentsline#1{#2}\egroup\vspace{-0.1in}}
\newcommand{\toclesslab}[3]{\vspace{0.1in}\bgroup\let\addcontentsline=\nocontentsline#1{#2\label{#3}}\egroup\vspace{-0.1in}}

\newcolumntype{P}[1]{>{\centering\arraybackslash}p{#1}}
\newcolumntype{C}{>{$}c<{$}} 

\begin{document}
\title{Learning Continuum-level Closures For Control of Interacting Active Particles}

\author{Titus Quah}
\affiliation{
Department of Chemical Engineering, University of California--Santa Barbara,
Engineering II Building, Santa Barbara, CA 93106
}
\author{Sho C. Takatori}
\affiliation{
Department of Chemical Engineering, Stanford University, 
443 Via Ortega, Stanford, CA 94305
}
\author{James B. Rawlings}
\affiliation{
Department of Chemical Engineering, University of California--Santa Barbara,
Engineering II Building, Santa Barbara, CA 93106
}

\date{\today}

\begin{abstract}
	Active matter swarms — collectives of self-propelled particles that could self-assemble, ferry microscopic cargo, or endow materials with dynamic properties — remain hard to steer. In crowded systems, tracking or controlling individual agents becomes challenging, so strategies should operate on macroscopic fields like particle density. Yet predicting how density evolves is difficult due to inter-agent interactions.
For model-based feedback control methods — like Model Predictive Control (MPC) — fast, accurate, and differentiable models are crucial. Detailed agent-based simulations are too slow, necessitating coarse-grained continuum models. However, constructing accurate closures — approximations expressing the effect of unresolved microscopic states (e.g., agent positions) on continuum dynamics in terms of the modeled continuum fields (e.g., density) — is challenging for active matter swarms.
We present a learning-for-control framework that learns continuum closures from agent simulations, demonstrated with active Brownian particles under a controllable external field. Our Universal Differential Equation (UDE) framework represents the continuum as an advection-diffusion equation. A neural operator learns the advection term, providing closure relations for microscopic effects like self-propulsion, interactions, and external field responses. This UDE approach, embedding universal function approximators in differential equations, ensures adherence to physical laws (e.g., conservation) while learning complex dynamics directly from data.
We embed this learned continuum model into MPC for precise agent simulation control. We demonstrate this framework's capabilities by dynamically exchanging particle densities between two groups, and simultaneously controlling particle density and mean flux to follow a prescribed sinusoidal profile. These results highlight the framework's potential to control complex active matter dynamics, foundational for programmable materials.
\end{abstract}
\maketitle

\toclesslab\section{Introduction}{sec:intro}

Active matter systems consist of agents that convert energy into directed motion, leading to a wide range of emergent phenomena, such as motility-induced phase separation (MIPS), flocking, and swarming \cite{marchetti:joanny:prost:rao:simha:2013,chate:2020,bechinger:dileonardo:reichhardt:volpe:volpe:2016,simha:ramaswamy:2002,ramaswamy:2010,sciortino:bausch:2021,vicsek:czirok:eshel:inon:ofer:1995,thampi:yeomans:2016,tailleur:cates:2008}.
Such phenomena promise transformative microrobotic capabilities (on-demand self-assembly, microscopic cargo transport, reconfigurable colloidal gels), yet steering them toward a specific, functional configuration remains notoriously difficult \cite{zhang:redford:bryant:gardel:depablo:2021,tayar:caballero:saleh:cristinamarchetti:dogic:2023,grober:palaia:hannezo:saric:palacci:2023,ramananarivo:ducrot:palacci:2019,zhou:wan:brady:sternberg:daraio:2024}.
To address this, various actuation mechanisms have been developed that harness external fields—magnetic, optical, acoustic, or electric—to influence agent behavior \cite{guillamat:ignesmullol:sagues:2016,palacci:sacanna:steinberg:pine:chaikin:2013,fernandezrodriguez:grillo:alvarez:rathlef:buttinoni:volpe:isa:2020,palagi:singh:fischer:2019,buttinoni:volpe:kummel:volpe:bechinger:2012,frangipane:dellarciprete:vizsnyiczai:bernardini:dileonardo:2018,lemma:varghese:thomson:baskaran:dogic:2023,arlt:martinez:dawson:pilizota:poon:2018,arlt:martinez:dawson:pilizota:poon:2019,takatori:dedier:vermant:brady:2016,mano:delfau:iwasawa:sano:2017,demirors:akan:poloni:studart:2018}.
Using particle-based control, researchers have successfully guided agents to form patterns, exhibit directed motion, and transport cargo \cite{franzl:cichos:2020,mano:delfau:iwasawa:sano:2017,pellicciotta:paoluzzi:frangipane:angelani:dileonardo:2023,massanacid:maggi:frangipane:dileonardo:2022,baldovin:gueryodelin:trizac:2023,koumakis:brown:griffiths:martinez:poon:2019,heuthe:panizon:gu:bechinger:2024,muinoslandin:fischer:holubec:cichos:2021,falk:alizadehyazdi:jaeger:murugan:2021,chennakesavalu:rotskoff:2021}. 
In active nematic systems, similar strategies have enabled flow speed control, defect steering, and the assembly and directed movement of asters \cite{nishiyama:berezney:hagan:dogic:fraden:2024,zhang:redford:bryant:gardel:depablo:2021,shankar:scharrer:bowick:marchetti:2024,ross:lee:banks:phillips:thomson:2019}. 
These methods often rely on the ability to track and influence individual agents/defects, a capability that becomes increasingly challenging, if not impossible, in semidense active systems. 
In such regimes, where individual particle identification and manipulation are lost, control strategies must instead focus on macroscopic measurables like density. This motivates the development of techniques that can shape and direct the collective behavior— i.e., active matter fields— by manipulating external fields to achieve desired density profiles and fluxes. 

Optimal control theory addresses this problem by formulating an optimal control problem, which involves predicting the system's response using a dynamic model over a time horizon and then optimizing the external field sequence to steer the system toward a desired outcome \cite{bertsekas:2017}. This approach successfully finds field sequences that steer active nematics, polar fluids, and droplets in prescribed manners, provided the model is accurate and disturbances are negligible \cite{norton:grover:hagan:fraden:2020,sinigaglia:braghin:serra:2024,davis:proesmans:fodor:2024,shankar:raju:mahadevan:2022,ghosh:baskaran:hagan:2025,ghosh:joshi:baskaran:hagan:2024}. However, real experimental settings rarely conform to these ideal conditions: fluctuations, and unmodeled interactions routinely nudge the system away from its intended path. In practice, then, feedback control—which updates the field based on new measurements—becomes essential for robustly managing unanticipated variations \cite{nishiyama:berezney:hagan:dogic:fraden:2024,franzl:cichos:2020,massanacid:maggi:frangipane:dileonardo:2022,mano:delfau:iwasawa:sano:2017}. Among optimal control based feedback strategies, one can employ model predictive control (MPC)—in which a finite horizon optimal control problem is iteratively solved after each new measurement, applying only the first field of the optimized sequence until the next measurement becomes available \cite{rawlings:mayne:diehl:2020,quah:modica:rawlings:takatori:2024c}. Alternatively, machine-learning approaches like reinforcement learning (RL), through trial and error, learn a policy that maps measured system states to the external fields to apply \cite{schulman:wolski:dhariwal:radford:klimov:2017,haarnoja:zhou:abbeel:levine:2018,ghosh:2025}. Both paradigms adapt to new measurements and disturbances, ensuring robust guidance toward the target behavior.

\begin{figure*}[t!]
	\centering
	\includegraphics[width=0.95\textwidth]{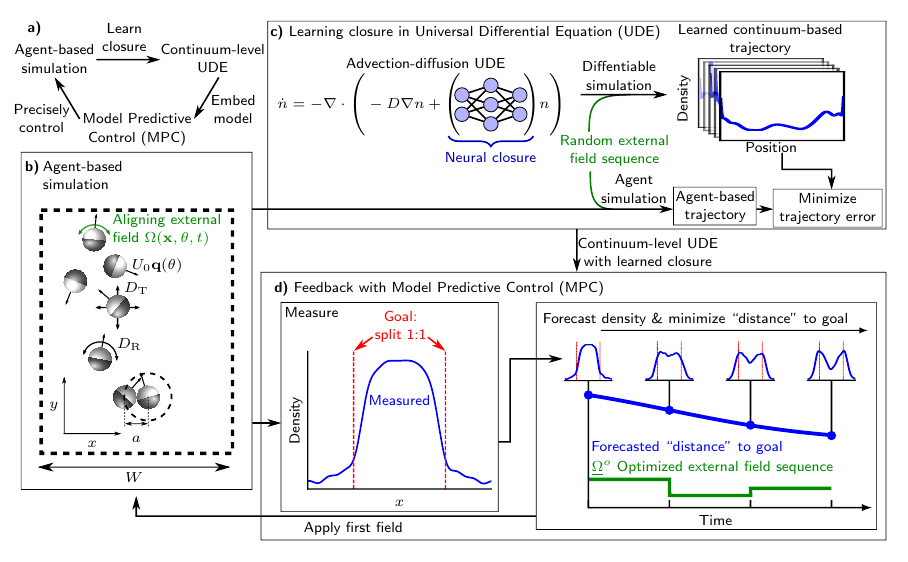}
	\caption{
	Scientific machine learning framework for modeling and control of interacting active matter.
	(\textbf{a}) Flow chart of the proposed framework. We use agent-based simulations to learn closures within a continuum-level universal differential equation (UDE). This UDE is then embedded into Model Predictive Control (MPC) to enable precise control over the agent-based simulation, guiding it towards desired behaviors.
    (\textbf{b}) Schematic of 2D interacting active Brownian particles (ABPs) under an actuated aligning external field that orients particles left or right.
    (\textbf{c}) The UDE is formulated as an advection-diffusion equation. A neural operator is used to represent the effective velocity field, learning its dependence on particle number density $n$, external field $\Omega$, and past densities and external fields $p$. The UDE is simulated using a differentiable solver under random external field sequences, starting from an initial density field. By comparing these predicted density trajectories with those from agent-based simulations, we learn the advection term, providing closure relations for microscopic effects like self-propulsion, interactions, and external field responses. Only macroscopic density fields, estimated from agent-based simulations, are used for training; individual particle trajectories are not directly used.
    (\textbf{d}) The learned UDE is embedded into MPC, enabling us to control emergent behaviors of the active matter system. MPC optimizes the external field sequence to achieve a desired goal, such as splitting the particle population into distinct groups or dynamically controlling particle fluxes. Our framework is robust, capable of handling particle interactions, and optimizes control actions for precise and adaptive manipulation of active matter.}
		   \label{fig:intro}
\end{figure*}

Whether using MPC or RL, both require a fast model of the density dynamics. While RL is often described as model-free, during training, RL still depends on extensive sampling of system trajectories, making fast model evaluations essential \cite{cichos:gustavsson:mehlig:volpe:2020,falk:alizadehyazdi:jaeger:murugan:2021,muinoslandin:fischer:holubec:cichos:2021}. MPC additionally requires differentiability so that gradients of the objective function with respect to the external field can be computed. While large-scale agent-based simulations faithfully capture microscopic details—such as self-propulsion, interactions, and responses to external fields—they are too slow to serve as models for either approach. Consequently, we pursue accurate, computationally efficient, and differentiable continuum-level simulations that effectively predict dynamic density responses to external fields. A central component in formulating such continuum-level models is developing closure relations for unresolved microscopic interactions and orientation fields, such as the polar field, expressed in terms of the density and external field \cite{hinch:leal:1976,ezhilan:shelley:saintillan:2013,weady:shelley:stein:2022}. Existing continuum models, including the active model B+, typically derive closures via the interaction expansion method and predominantly examine steady-state scenarios or minor spatial and temporal variations in external fields \cite{wittkowski:tiribocchi:allen:marenduzzo:cates:2014,vrugt:bickmann:wittkowski:2023}. Although effective in replicating steady-state phenomena like MIPS, our focus extends to quantitatively capturing dynamic behaviors under significant and rapid external field variations.

Instead of manually deriving the closure in the continuum-level model, we propose a scientific machine learning framework that leverages agent-level simulation data to learn the necessary closure relations. This learned continuum-level model is then incorporated into MPC, enabling precise control of agent-level simulations to achieve prescribed trajectories \cref{fig:intro}a. We demonstrate our proposed framework using a system of $10^4$ interacting active Brownian particles (ABPs) subject to an  actuated external aligning field, as illustrated in \cref{fig:intro}b. To effectively capture the density dynamics, we represent the continuum model as an advection-diffusion equation and employ a neural operator to close the advection term, accounting for influences from the external field, self-propulsion, and multibody interactions (\cref{fig:intro}c). Such formulations belong to the class of \emph{universal differential equations} (UDEs), which embed universal function approximators, like neural networks, within differential equations \cite{rackauckas:ma:skinner:ramadhan:edelman:2021}. Embedding these approximators within differential equations enforces ``physics'', e.g. conservation laws or symmetries, while providing the flexibility to accurately learn complex dynamics from limited data. This approach not only reduces the required training dataset compared to purely data-driven models but also ensures adherence to essential physical laws \cite{faroughi:pawar:das:kalantari:kouroshmahjour:2024}. Similar methods have previously demonstrated efficacy in discovering constitutive models from experimental data in viscoelastic materials, and for closure modeling in active fluids and coarse-grained fluid dynamics \cite{lennon:mckinley:swan:2023,maddu:weady:shelley:2024,debezenac:pajot:gallinari:2019,sanderse:stinis:maulik:ahmed:2024,srivastava:duraisamy:2021}. Nevertheless, an accurate continuum model alone does not guarantee robust controller performance when embedded in a control strategy \cite{kumar:rawlings:2023a}.

To validate the continuum model's effectiveness within an optimal control framework, we integrate the continuum model into MPC and perform two illustrative control experiments. In the first scenario, we partition the particle population into two distinct groups and dynamically exchange particle densities between them, highlighting the framework's capability for precise spatial and temporal manipulation of number densities (\cref{fig:intro}~d). The second experiment demonstrates simultaneous control of both particle density and mean flux, guiding the latter to follow a predefined sinusoidal trajectory. In this case, the controller modulates particle alignment and external fields, regulating particle mean flux while maintaining particle groups concentrated around their intended positions. These control scenarios underscore the learning-for-control framework's potential to accurately learn intricate dynamics and effectively optimize control actions, offering significant promise for expanded applications within active matter systems.

\toclesslab\section{Learning the advection-diffusion UDE from agent-based simulations}{sec:learn_ude}
\begin{figure*}[t!]
	\centering
\includegraphics[width=0.95\textwidth]{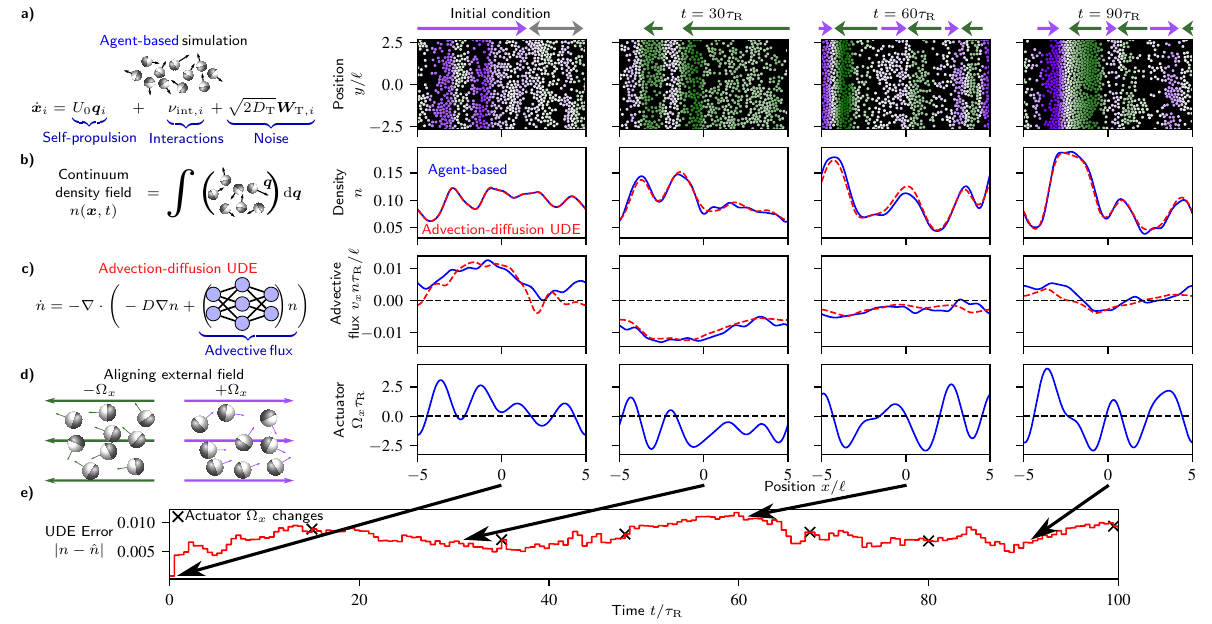}
\caption{Prediction accuracy of the UDE over a 100 $\taur$ time horizon under random external field changes. (\textbf{a}) Agent-based simulation snapshots for the initial condition and three time points: $t=[30\taur, 60\taur, 90\taur]$. Particles are colored based on their polar order in the $x$ direction, normalized by density, $m_x(x)/n(x)$, where $m_x(x) = \int\int \cos(\theta)P(\bm{x}, \theta, t),\mathrm{d}\theta,\mathrm{d}y$. Positive polar order (particles oriented to the right) is shown in purple, while negative polar order (particles oriented to the left) is shown in green. Regions without strong polar order are colored white. Colored arrows above the particles are included for visual clarity. (\textbf{b}) Number densities are obtained by integrating out orientation $\bm{q}$. The initial agent-based (blue) density initializes the advection-diffusion UDE simulation (red), which is simulated forward for 100$\taur$ with varying external fields. Comparisons between the agent-based and UDE simulations are shown for the same time points as in (\textbf{a}). (\textbf{c}) The advective flux due to the velocity field $v_xn$ is displayed for agent-based (blue) and UDE (red) simulations, with fluctuations attributed to particle collisions. (\textbf{e}) The external field applied during the simulation is shown for context. (\textbf{f}) The $L^2$ norm of the error between the predicted and measured number densities remains below 0.01 throughout the simulation, indicating a close match to the true number density profile. X-marks indicate points where the external field changes.}
\label{fig:error}
\end{figure*}

We consider a system of $N$ active Brownian particles (ABPs) in two spatial dimensions as shown in \cref{fig:intro}~b. Each particle, located at position $\bm{x}$ with orientation angle $\theta$, self-propels with speed $U_0$ in direction $\bm{q} = \begin{bmatrix}\cos\theta&\sin\theta\end{bmatrix}^\mathrm{T}$. They interact via short-range repulsive forces, mimicking hard-core exclusion, characterized by particle radius $a$. Additionally, particles undergo translational and rotational diffusion with coefficients $D_\mathrm{T}$ and $D_\mathrm{R}$, respectively. Their orientations can be manipulated by a spatiotemporally varying external field, characterized by its strength $\Omega_x(x,t)$, which induces an angular velocity $\Omega$ that aligns particles left or right. The external field has a maximum strength $\Omega_{\max}$. Throughout this work, we use the rotational diffusion time, $\taur = 1/D_\mathrm{R}$, and the run length, $\ell = U_0 \taur$, as characteristic units for time and length, respectively. The full simulation details can be found in \cref{sec:bd}.

We focus on controlling the one-dimensional $x$-number density, $n(x,t)$, a key macroscopic measurable of these dense active matter swarms. The $x$-number density $n(x,t)$ is defined as the the one-particle distribution function $P(\bm{x},\theta,t)$ marginalized over orientation $\theta$ and spatial dimension $y$:
\begin{equation}
n(x,t) = \int \int P(\bm{x},\theta,t) \,\mathrm{d}\theta\, \mathrm{d}y
\label{eq:number_density}
\end{equation}
$x$ and $y$ are the spatial dimensions; $\bm{x}=\begin{bmatrix}x&y\end{bmatrix}^T$; $\theta$ is the orientation angle.

Our control strategy employs MPC to determine an optimal sequence of external fields $\bm{\Omega}^o$ that drives the number density $n(x,t)$ toward a desired goal as illustrated in \cref{fig:intro}~d. Anticipating that system perturbations and model uncertainties will cause deviations between the intended and observed trajectories, we incorporate feedback control to adjust the external field accordingly. Our implementation operates under the constraint that feedback is available only through measurements of the number density $n(x,t)$, taken at sample time intervals $h$.

However, to proceed with MPC, we first require a predictive model for the number density dynamics $n(x,t)$. As discussed in \cref{sec:smol}, deriving accurate closures for continuum-level models is challenging, so instead, we formulate the density dynamics model as an advection diffusion UDE, closing the equations with a neural operator that learns the advection term as follows:
\begin{gather}
	\pder[\hat{n}(x,t)]{t} + \pder{x}\left(-D_\mathrm{T} \pder{x}\hat{n}(x,t)+\hat{v}_x(x,t)\hat{n}(x,t)\right) = 0
	\label{eq:gray-box-n}\\
	\hat{v}_x(x,t) =v_\mathrm{NN}(\hat{n}(x,t), \Omega_x(x,t), p(x,t);\beta_\mathrm{NN})
\end{gather}
where $\hat{n}(x,t)$ and $\hat{v}_x(x,t)$ are the predicted number density and velocity fields, respectively; The neural operator $v_\mathrm{NN}(\cdot)$ approximates the effective velocity field; $p(x,t)$ contains $N_p$ past number density and external fields, and $\beta_\mathrm{NN}$ representing the neural operator parameters. Since we only model number density dynamics without access to higher-order moments (e.g., polar order or the two-particle distribution function), we incorporate past densities and external fields $p(x,t)$ to approximate these unmeasured states, following Taken's embedding theorem \cite{takens:1981}. For non-interacting ABPs (a linear system), one can show that embedding $N_p$ past density fields is equivalent to including $N_p$ moments in the moments expansion method \cite{takatori:quah:rawlings:2025}. 
Through the advection-diffusion structure, this formulation ensures mass conservation and non-negative densities. 

Given our discrete-time measurements with interval $h$, we hold the external field constant between each measurement time and integrate \cref{eq:gray-box-n} using spectral methods and method of lines with time step $h$ to derive the discrete-time dynamics $f_h(\cdot)$:

\begin{equation}
	\hat{n}(k+1) = f_h(\hat{n}(k),\Omega_x(k),v_\mathrm{NN}(\hat{n}(k), \Omega_x(k), p(k);\beta_\mathrm{NN}))
\end{equation}
$k$ denotes the discrete time index. We omit the spatial dependence in $x$ for brevity.
To determine the neural operator parameters $\beta_\mathrm{NN}$, we collect number density trajectories by applying random external field sequences following the protocol in \cref{sec:train_d}. Briefly, we performed 2D periodic BD simulations of interacting ABPs, applying random spatial collocation of external fields in 1D over randomly selected time durations. The $x$-number density is represented in a spectral basis and estimated from particle positions through orthogonal series estimation, as detailed in \cref{sec:orth} \cite{cencov:1962}. To capture both transient and steady-state dynamics, the external field was held constant over random intervals, allowing the ABPs to sample diverse configurations. Using the initial frames to set $\hat{n}(x,0)$ and $p(x,0)$, we generate predicted trajectories and optimize $\beta_\mathrm{NN}$ to minimize the discrepancy between agent-based and UDE-predicted trajectories, as illustrated in \cref{fig:intro}~c. 

We set the sampling time to $h=0.5\taur$ and utilize $N_p=2$ past fields. Our dataset spans $10^4\taur$, encompassing nearly 2000 unique external fields. The implementation leverages PyTorch for the neural operator architecture, employs a modified Backward Euler method for discrete-time dynamics, and utilizes the Adam optimizer for loss minimization. Since the density field dynamics evolve on a slower time scale than the discrete Langevin particle dynamics, we are able to take large time steps in our implicit solver. Detailed training protocols are documented in \cref{sec:learn_flux}. 
\toclesslab\subsection{UDE prediction accuracy}{sec:val}
\cref{fig:error} shows the prediction accuracy of the UDE over 100 $\taur$ with an initial condition and external field sequence not used during training, i.e., testing data. A video of the test data comparison is shown in Movie S1. \cref{fig:error}~a-d show snapshots of the test trajectory with the first column showing the initial condition and columns [2-4] corresponding to $t=[30\taur, 60\taur, 90\taur]$, respectively. \cref{fig:error}~a shows snapshots of the particle simulation. Note the entire simulation has a height of about $50\ell$ so we show only a subsection. Particles are colored to indicate their orientation: purple for right-oriented particles and green for left-oriented particles. Orientation is determined by the normalized polar order in $x$, defined as $m_x(x)/n(x)$, where $m_x(x) = \int \int \cos(\theta) P(\bm{x}, \theta, t) \, \mathrm{d}\theta \, \mathrm{d}y$. Positive values of $m_x(x)/n(x)$ correspond to right-oriented particles (purple), while negative values correspond to left-oriented particles (green).
Regions without strong polar order are colored white. 
Colored arrows above are to guide the eye. 
\cref{fig:error}~b shows agent-based simulation (blue) and UDE-predicted (red) snapshots of the density $n$. \cref{fig:error}~c shows the agent simulation (blue) and UDE-predicted (red) flux due to velocity field $v_xn$. 
We emphasize that the UDE learns the effective velocity $v_x$ based solely on the density field dynamics $n(x,t)$ as input data, rather than directly using the measured values of $v_x$.
\cref{fig:error}~d shows the external field at the time of the snapshot. \cref{fig:error}~e shows the $L^2$ norm of the error between the UDE-predicted and measured number densities with x-marks marking when the external field changes.

As can be seen in \cref{fig:error}~e, the error norm remains below 0.01 and this corresponds to a number density profile that closely resembles the measured density profile as seen in \cref{fig:error}~b. 
Furthermore, there are no sudden increases in the error norm when the external field changes, suggesting the dynamics are well captured. Crucially, this 100 $\taur$ UDE simulation takes only $\sim\!0.1$\,s to evaluate, roughly $10^{4}\times$ faster than the agent-based simulation ($\sim\!10^{3}$\,s).

 Turning our attention to the advective flux in \cref{fig:error}~c, despite never explicitly providing the measured flux to the UDE, we can see by matching the number density, the UDE also matches the advective flux throughout the test simulation, minus the fluctuations induced by the multibody particle collisions and thermal noise. These results demonstrate that the UDE accurately captures both the number density and emergent flux behavior, validating its suitability for implementation within the MPC framework.

\toclesslab\section{Model Predictive Control (MPC)}{sec:mpc}
\begin{figure*}[t!]
	\centering
\includegraphics[width=0.95\textwidth]{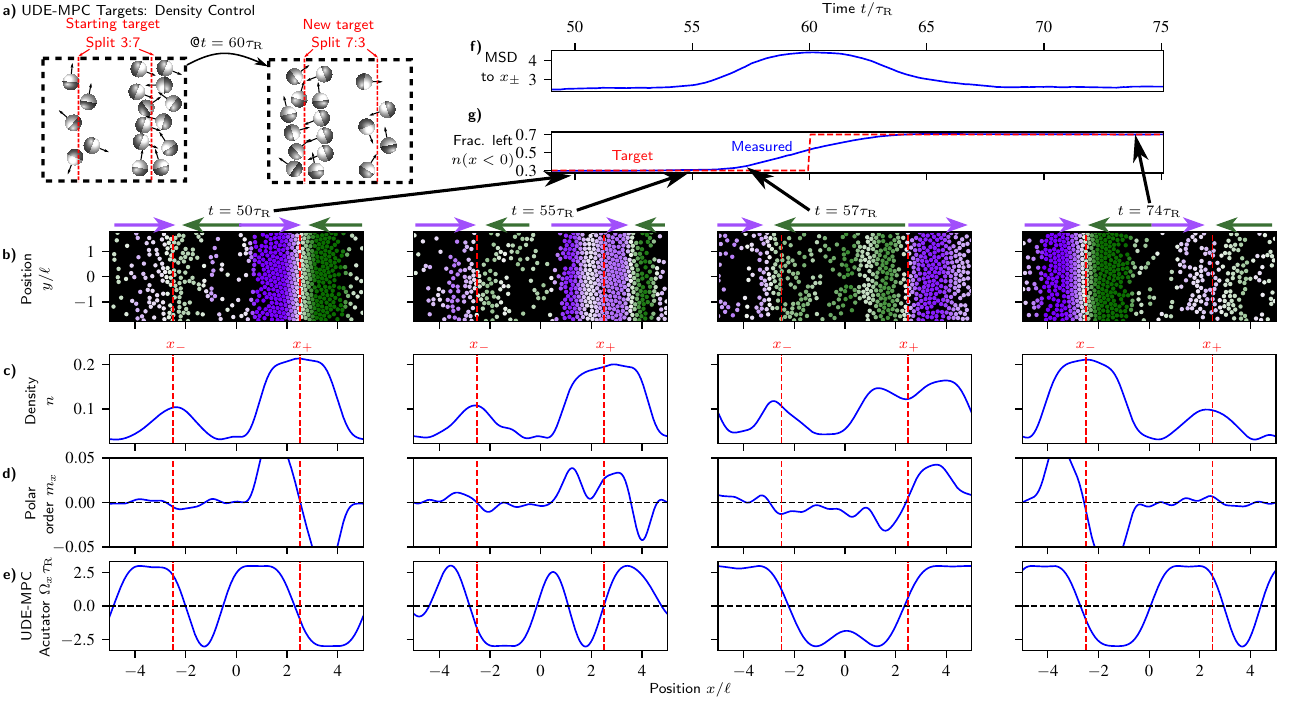}
\caption{Density control of agent-based simulations using UDE-MPC. (\textbf{a}) UDE-MPC targets are to first ``trap'' particles at the targets $x_{+}$ and $x_{-}$ with a left-to-right density ratio of 3:7. At $t=60\taur$, the target left-to-right density ratio changes to 7:3 with the targets remaining at $x_{+}$ and $x_{-}$. UDE-MPC has a control horizon of $7\taur$ and begins planning for upcoming target change at $t=53\taur$. (\textbf{b}) Snapshots of the particle simulation at four key time points: $t = 50\taur$, $55\taur$, $57\taur$, and $74\taur$. Particles are colored based on their expected orientation in the $x$ direction, averaged over the $y$ direction: purple indicates particles oriented to the right, green indicates particles oriented to the left, and white indicates no strong orientation. Red dashed lines mark the targets $x_{+}$ and $x_{-}$. Colored arrows above the particles guide the eye. (\textbf{c}) Snapshots of the number density field at the same four time points. (\textbf{d}) Snapshots of the polar order field. (\textbf{e}) The external field applied to the system obtained from UDE-MPC, is displayed. We refer to points where the external field switches from positive to negative as attractive wells and locations where the field switches from negative to positive as repulsive hills. Notice that in preparation for the ratio target change at $t=60\taur$, the controller begins planning at $t=53\taur$ and takes action by orienting particles in the $x_{+}$ group leftward (green) while jamming particles near the $x_{+}$ target to maintain proximity. This behavior creates a pressure front, preparing the particles for transfer to $x_{-}$ while preventing excessive displacement from $x_{+}$.  (\textbf{f}) The sum of the mean squared distances (MSDs) of particles to their respective targets, $x_{+}$ and $x_{-}$, is plotted. (\textbf{g}) The target (red line) vs. agent-based simulation (blue line) density fraction of particles to the left of the origin.
    }
 \label{fig:dance}
\end{figure*}
MPC is a feedback control strategy that uses a system's dynamic model to predict its future behavior over a short time horizon. At each step, it calculates the optimal control actions to achieve a desired goal, while respecting physical limitations (e.g., the external field's maximum strength). Crucially, only the first action from this calculated sequence is applied, and the entire planning process is then re-evaluated and updated with new measurements (\cref{fig:intro}~d). Formally, provided initial conditions $n_0$ and $p_0$,
we solve the following nonlinear optimization problem to find the optimal control sequence $\underline{\Omega}^\mathrm{o}$ over a control time horizon $K$, minimizing over the sequence of actions, $\underline{\Omega}$:
\begin{mini}|s|
	{\underline{\Omega}}{\sum_{k=0}^{K-1}L(\hat{n}(k), \underline{\Omega}(k)) + L_f(\hat{n}(K))}
	{}{}
	\addConstraint{\hat{n}(k+1) = }{f_h(\hat{n}(k),\underline{\Omega}(k),v_\mathrm{NN}(\hat{n}(k), \underline{\Omega}(k), \hat{p}(k);\beta_\mathrm{NN}))}
	\addConstraint{\hat{p}(k+1) = }{f_p(\hat{n}(k),\underline{\Omega}(k),\hat{p}(k))}
	\addConstraint{\hat{n}(0) =}{n_0}
	\addConstraint{\hat{p}(0) =}{p_0}
	\addConstraint{\underline{\Omega}(k)\subseteq\mathbb{W}\quad \forall k\in[0,K-1]}
	\label{eq:mpc}
\end{mini}
 $\underline{\Omega}(k)$ is the $k$th element of the sequence $\underline{\Omega}$; $L(\cdot)$ is the stage cost and $L_f(\cdot)$ is the terminal cost; $f_p(\cdot)$ is a sliding window function that drops the oldest element of the past vector $\hat{p}(k)$ and appends the newest elements, namely, $\hat{n}(k)$ and $\underline{\Omega}(k)$;
The stage cost is typically chosen to penalize the deviation of the number density from the set point $n_{\mathrm{sp}}(x,t)$ and the control effort; The set $\mathbb{W}$ defines the admissible control actions, incorporating practical limits on the external field, such as their maximum strength.
As discussed, intended trajectories often deviate from observed ones due to model errors or system perturbations, necessitating feedback to correct the external field sequence. MPC inherently provides this by iteratively solving the optimization problem at each sample time $h\taur$, initializing the problem with the latest measurements, and applying only the first element of the optimal control sequence, as depicted in \cref{fig:intro}~d.

We demonstrate the UDE-MPC framework using two examples. In the first, we split the population into two groups and dynamically juggle the densities between them, showcasing precise control over spatial distributions. In the second, we simultaneously regulate the particle density and the mean flux, guiding the mean flux to follow a prescribed sinusoidal profile while maintaining the particle groups at their target positions. For both examples, we consider a control horizon $K=14$, i.e., a control horizon of $7\taur$, and a sample time $h=0.5\taur$.

The splitting example is shown in Movie S2, and the simultaneous control of density and mean flux is demonstrated in Movie S3. Additional examples are provided to highlight the framework's versatility. In Movie S4, we again show the results for Movie S2, but include the MPC forecast. In Movie S5, we revisit the splitting example but reformulate the stage cost using the Wasserstein distance, illustrating how the framework can fit particles to arbitrary distributions \cite{villani:2009}. In Movie S6, we design the fastest trap of width $4\ell$ moving to the right that keeps 60\% of the particles in the trap. Code is available at \url{https://github.com/titusswsquah/graybox_abp_mpc} \cite{quah:takatori:rawlings:2025a}.

\toclesslab\section{Splitting and juggling the population}{sec:splitties}
For this example, we initialize the simulation with the particles accumulated at the center. 
The user targets are as follows:
\begin{enumerate}
	\item Split particles into two equal groups centered at $x_{+}=2.5\ell$ and $x_{-}=-2.5\ell$, where $\ell = U_0 \tau_\text{R}$ is the run length.
	\item Achieve a distribution where the probability of finding a particle in the left half (i.e., with position $x<0$) is 30\% while maintaining the populations centered at $x_{\pm}$.
	\item Repeat step 2, but with the probability of finding a particle in the right half is 70\%.
	\item Return to step 1.
\end{enumerate}
A movie of the full process is shown in Movie S2. The stage cost for this example penalizes three terms: mean squared distance (MSD) of particles in the left half to the left target $x_-$ and similarly MSD of particles in the right half to the right target $x_+$, and a term that penalizes the difference between the target and agent-based simulation left-to-right density ratio.
We defer the formal stage cost definition to \cref{sec:stg_cost1}.

\begin{figure*}[t]
	\centering
\includegraphics[width=0.95\textwidth]{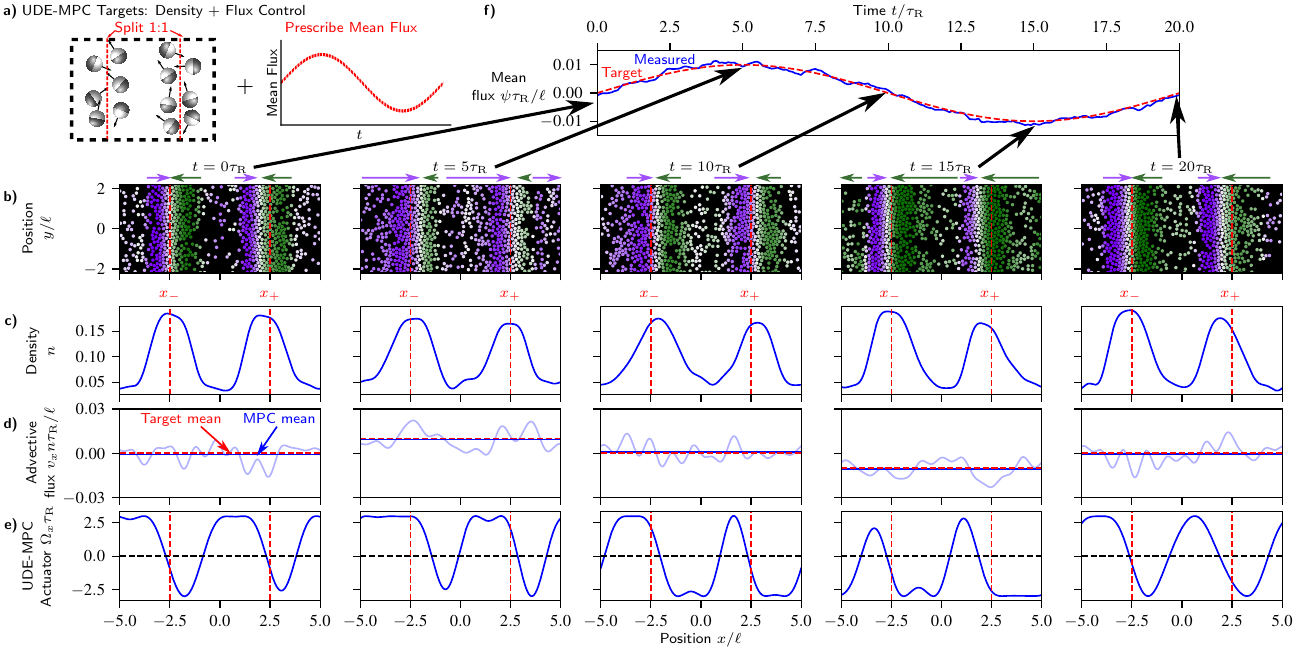}
\caption{
	Simultaneous control of particle number density and mean flux using UDE-MPC. (a) UDE-MPC targets are to simultaneously maintain density accumulation about targets $x_+$ and $x_-$ and have the mean flux over all the particles follow a sinusoidal profile in time (b) Particle positions and polar order at representative times, highlighting leftward (green) and rightward (purple) orientations. (c) Density profiles consistently accumulate near $x_\pm$, indicating the successful maintenance of the target accumulation. (d) Particle advective flux profiles confirm precise mean flux control, with measured mean flux (solid blue line) matching the set point (dashed red line). (e) external field snapshots reveal dynamic control strategies, including field modulation to achieve target fluxes, paired with localized jamming around $x_\pm$ to maintain density accumulation about $x_\pm$. Red dashed vertical lines indicate target particle accumulation points, emphasizing the controller's ability to simultaneously regulate both particle density and flux across the domain. (f) The set point particle mean flux (dashed red line) and measured flux (solid blue line) show accurate tracking throughout the sinusoidal profile.
	}
	\label{fig:force_ctrl}
\end{figure*}

\toclesslab\subsection{Controller behavior}{sec:ctrl2}

We examine the transition from step 2 to step 3, focusing on the transfer of particles from the $x_{+}$ to the $x_{-}$ group as shown in \cref{fig:dance}. The set point change occurs at $t=60\taur$, and we analyze the controller's behavior during the time interval $[50\taur, 75\taur]$.
\cref{fig:dance}~a illustrates the control objective which is to transition from a left-to-right density ratio of 3:7 to 7:3 while keeping particles accumulated about $x_{\pm}$. The lower panels present simulation snapshots at four time points: $t = 50\taur$, $55\taur$, $57\taur$, and $74\taur$.

\cref{fig:dance}~b illustrates the particle simulation snapshots, with particles colored according to their expected orientation along the $x$-axis, averaged over the $y$ direction. Purple indicates rightward orientation, green indicates leftward orientation, and white indicates no strong directional preference. Guiding arrows above the particles highlight these orientations, while red dashed lines mark the set points $x_{\pm}$.
The remaining panels show complementary measurements: \cref{fig:dance}~c presents the density field, \cref{fig:dance}~d displays the polar order, and \cref{fig:dance}~e shows the external field applied to the system, obtained from UDE-MPC. 
\cref{fig:dance}~f displays the combined costs of maintaining populations at $x_{\pm}$, represented by the sum of mean squared distances between the left (right) particles and $x_{-}$ ($x_{+}$). In \cref{fig:dance}~g, we compare the target fraction (red) and measured fraction of particles to the left of $x=0$.

At the beginning of the interval, particles are distributed around their respective set points, with the left particles clustered near $x_{-}$ and right particles near $x_{+}$. The simulation snapshot at $t=50\taur$ in \cref{fig:dance}~b and the corresponding density profile in \cref{fig:dance}~c show a smaller population near the left set point, maintaining the target fraction of $n(x<0)\approx 0.3$.
The controller's behavior, illustrated in \cref{fig:dance}~e, directs the particles toward the set points.  Specifically, the field transitions from positive to negative values, creating ``attractive wells'' near the set points. Conversely, transitions from negative to positive values generate ``repulsive hills'', which appear at the simulation cell boundaries ($x=\pm 5\ell$) and at the center ($x=0$). These repulsive regions effectively channel particles away from these locations and toward the designated set points.
The effectiveness of this orienting field is evident in \cref{fig:dance}~d, where the polar order parameters exhibit transitions from positive to negative values in the vicinity of $x_{\pm}$, confirming the controlled directional behavior of the particles.

Recall we are using a control horizon of $7\taur$, so for the upcoming set point change occurring at $t=60\taur$, the controller begins planning at $t=53\taur$. At this point, the controller initiates action through a sophisticated approach that leverages two fundamental features of this system: the limited external field and jamming dynamics. Limited by the maximum external field magnitude $\Omega_{\max}$, the controller cannot instantaneously reorient particles from the right group to the left group. Instead, from $t=53\taur$ to $t=56\taur$, the controller methodically prepares the particles by orienting the majority of particles in the $x_+$ group toward the left, as evidenced by their green coloration in the $t=55\taur$ simulation snapshot (\cref{fig:dance}~b).
This reorientation, however, presents a challenge: the particles' self-propulsion causes them to move rightward once oriented in that direction. The controller addresses this through a strategic jamming mechanism, directing a layer of particles (displayed in purple) to the left of the group to move rightward, thereby opposing the main group's flux. This creates an effective pressure front that serves two purposes: preparing the particles for their leftward migration while maintaining their position near the $x_+$ set point.
This preparation phase is apparent in both the polar order (\cref{fig:dance}~d) and the external field (\cref{fig:dance}~e). The controller maintains an attractive well at $x_-$ while establishing a more complex structure around $x_+$: a repulsive hill flanked by two attractive wells that effectively jam the particles to prevent escape. The polar order reveals a transition from positive to negative values to the left of the $x_+$ set point. Notably, the number density shown in \cref{fig:dance}~c remains relatively unchanged compared to the $t=50\taur$ snapshot, a consistency that is reflected in the minimal variations observed in both the sum of MSDs (\cref{fig:dance}~f) and the left fraction (\cref{fig:dance}~g).

From $t=56\taur$ to $t=61\taur$, the controller floods particles leftward, transferring them from $x_+$ to $x_-$ as shown in the $t=57\taur$ snapshot (\cref{fig:dance}~b,c). This transfer is driven by an attractive well at $x_-$ and a repulsive hill at $x_+$ in both the polar order and external field (\cref{fig:dance}~d,e). During this period, the sum of MSDs increases, peaking at $t=60\taur$ (\cref{fig:dance}~f), while the left fraction approaches its 70\% set point (\cref{fig:dance}~g).
From $t=61\taur$ to $t=75\taur$, the controller directs particles to their respective set points, maintaining attractive wells at both $x_-$ and $x_+$ (\cref{fig:dance}~d,e). The left fraction reaches its 70\% set point by $t=64\taur$ (\cref{fig:dance}~g), and the controller continues minimizing the sum of MSDs until $t=69\taur$ (\cref{fig:dance}~f).

\toclesslab\section{Advective flux control}{sec:adv}
In the previous example, we focused solely on controlling the number density. Given the UDE's accurate prediction of advective flux, we now extend our approach to simultaneously control both the advective flux and number density. We maintain the two-population particle system accumulated near $x_\pm$ while introducing control of the system's mean flux $\psi$ to follow a sinusoidal profile as shown in \cref{fig:force_ctrl}~a. Our stage cost is similar to the population splitting example, but with an additional penalty for differences between the target and UDE-predicted mean fluxes (see \cref{sec:stg_cost2} for details).

 A full process visualization is provided in Movie S3, while key snapshots are presented in \cref{fig:force_ctrl}. Panels (b)-(e) illustrate simulation snapshots of particle positions (with green/purple arrows indicating leftward/rightward polar order), density, flux (dashed red for the set point, solid blue for measured mean, and blue shaded line for field), and the external field. Red dashed vertical lines mark target particle accumulation positions. The set point flux (dashed red line) and measured flux (solid blue line) are shown in panel (f).

At the simulation's start ($t=0$), the two populations are equally distributed about $x_\pm$, oriented roughly towards $x_\pm$, as depicted by the particles and their colors in \cref{fig:force_ctrl}~b. This configuration corresponds to density peaks at $x_\pm$ in \cref{fig:force_ctrl}~c. The initial set point mean flux is zero, and UDE-MPC successfully achieves this flux, as evidenced by the dashed red and solid blue lines overlapping in \cref{fig:force_ctrl}~d. The external field initially behaves as it does when splitting the populations, creating two attractive wells at $x_\pm$. The repulsive hills are shifted left to start directing particles to the right, as the target flux begins to increase.

By $t=5\taur$, the set point mean flux reaches its maximum value of 0.01. UDE-MPC accurately tracks the set point, as shown in \cref{fig:force_ctrl}~d. To achieve this, the controller directs most particles to the right, indicated by the purple particles in \cref{fig:force_ctrl}~b and the positive external field over most of the domain in \cref{fig:force_ctrl}~e. Despite this flux, the density remains accumulated near $x_\pm$, as seen in \cref{fig:force_ctrl}~c. To maintain density accumulation at $x_\pm$, the controller creates a jam at $x_\pm$ by redirecting particles to the right of $x_\pm$ leftwards. This behavior ensures that particles stay localized near $x_\pm$ while the desired mean flux is achieved. This is evident from the green particles on the right of $x_\pm$ in \cref{fig:force_ctrl}~b and the negative external field to the right of $x_\pm$ in \cref{fig:force_ctrl}~e.

The simulation follows a similar pattern in the opposite direction. At $t=10\taur$, the particles regroup near $x_\pm$ with zero mean flux. At $t=15\taur$, the set point flux reaches its minimum value of -0.01. UDE-MPC achieves this flux by applying a predominantly negative field across the domain, directing particles leftward. Again, to jam the particles about $x_\pm$, the controller creates an opposing group to the left of $x_\pm$, directing them rightward with a positive field, resulting in the desired mean flux. Throughout, particles remain concentrated near $x_\pm$. Finally, at $t=20\taur$, the simulation concludes with particles gathered near $x_\pm$ and zero mean flux. Overall, UDE-MPC successfully controls the agent-based simulation's agent mean flux to follow the prescribed sinusoidal profile as seen in \cref{fig:force_ctrl}~f. This strategy demonstrates the ability to simultaneously control number density and advective flux.

\toclesslab\section{Discussion}{sec:disc} 

In this work, we have developed and demonstrated a UDE-MPC framework for active matter systems, focusing on interacting ABPs. By representing the continuum dynamics with an advection-diffusion UDE, we accurately learned the dynamics of the number density from agent-based simulations and integrated this predictive model into an MPC framework to control emergent behaviors. We validated the framework with two illustrative examples: first, demonstrating precise spatial and temporal density control by dynamically exchanging particle populations between two groups; and second, achieving simultaneous control of mean particle flux, guiding it along a sinusoidal trajectory, while concurrently maintaining particle density within concentrated regions.  These results highlight the framework's ability to handle complex, multi-objective control tasks in active matter systems, demonstrating both accuracy and flexibility.

While our focus in this work was on learning the advective operator, there are natural extensions for cases where learning the diffusion operator is also critical. For example, in the Stokes limit, where hydrodynamics are governed by Stokes equations, particles exhibit long-range interactions due to flow fields induced by their motion. In such systems, the diffusion operator becomes a mobility tensor that depends on the system's configuration, requiring a closure relation that accounts for the number density, external fields, and their histories \cite{elfring:brady:2022}. Similarly, the fluctuations in the density field motivates exploring stochastic partial differential equations (PDEs). This exploration can build upon existing work on stochastic physics-informed neural ordinary differential equations (ODEs) \cite{oleary:paulson:mesbah:2022} by extending their methodology to PDEs through established numerical techniques like spectral methods and the method of lines. These techniques enable the transformation of PDEs into systems of ODEs, allowing us to learn the moments of the number density and velocity while leveraging the original framework's capabilities for handling stochastic elements.

Looking beyond this study, our framework has the potential to impact experimental systems, such as light-activated bacteria or other active colloidal systems, where bacteria density dynamics can be learned and controlled \cite{frangipane:dellarciprete:vizsnyiczai:bernardini:dileonardo:2018,arlt:martinez:dawson:pilizota:poon:2018}. Additionally, the framework's focus on learning dynamics and closures from data suggests applications in entirely different domains, such as improving continuum traffic models for urban planning and vehicle coordination \cite{patsatzis:russo:kevrekidis:siettos:2023}.

Ultimately, this work highlights the promise of combining scientific machine learning with MPC to tackle the complexity of active matter systems. By bridging the gap between data-driven methods and physical principles, our approach addresses a long-standing challenge in the field: systematic control of emergent behaviors. With its ability to adapt to disturbances and optimize multi-objective tasks, this framework represents a scalable and versatile tool for future research and applications in active matter and beyond.

\toclesslab\section{Acknowledgments}{sec:acknow}
We would like to thank Steven J. Kuntz and Sachit G. Nagella for helpful discussions.
T.Q. is supported by the National Science Foundation Graduate Research Fellowship Program under Grant No.~2139319. 
S.C.T. is supported by the Packard Fellowship in Science and Engineering.
Use was made of computational facilities purchased with funds from the National Science Foundation (CNS-1725797) and administered by the Center for Scientific Computing (CSC). The CSC is supported by the California NanoSystems Institute and the Materials Research Science and Engineering Center (MRSEC; NSF DMR 2308708) at UC Santa Barbara. 

\vspace{0.25cm}
\noindent
\textbf{Code availability.}
Code is available at \url{https://github.com/titusswsquah/graybox_abp_mpc} \cite{quah:takatori:rawlings:2025a}.

\onecolumngrid
\newpage
\centerline{\large\textbf{Supplementary Information}}
\setcounter{page}{1}
\appendix
\renewcommand\thefigure{S\arabic{figure}}    
\setcounter{figure}{0}

\tableofcontents

\section{Methods}
\subsection{Brownian Dynamics Simulation}\label{sec:bd}
We consider a system of $N$ interacting ABPs in two spatial dimensions as shown in \cref{fig:intro}~b.
In addition, the particles' orientations can be manipulated via a spatiotemporally-varying external field that induces an angular velocity. 
The dynamics of particle $i$ with center of mass position $\bm{x}_i = \begin{bmatrix}x_i&y_i\end{bmatrix}^\mathrm{T}$, orientation $\bm{q}_i = \begin{bmatrix}\cos(\theta_i)&\sin(\theta_i)\end{bmatrix}^\mathrm{T}$ and orientation polar angle $\theta_i$, where $i = 1,\ldots N$, can be described by the following Brownian dynamics equations, where the overdot denotes a time derivative:

\begin{gather}
	\dot{\bm{x}}_i = U_0\bm{q}_i + \nu_{\mathrm{int},i} + \sqrt{2D_\mathrm{T}}\bm{W}_{\mathrm{T},i}(t)
 \label{eq:stoc_gov1}
 \\
	\dot{\theta}_i = \Omega_i + \sqrt{2D_\mathrm{R}}W_{\mathrm{R},i}(t),
\label{eq:stoc_gov2}
\end{gather}
Here, $U_0$ represents the self-propulsive speed; the interaction term $\nu_{\mathrm{int},i}$ takes the form $\nu_{\mathrm{int},i}=- (D_\mathrm{T}/k_\mathrm{B}T)  \sum_{i=1,i\neq j}^N\nabla_{i} U_2(\norm{\bm{x}_i-\bm{x}_j})$; $D_\mathrm{T}$ is the translational diffusion coefficient; $k_\mathrm{B}T$ is the thermal energy scale; 
The gradient operator with respect to $\bm{x}_i$ is denoted as $\nabla_{i}= \begin{bmatrix}\pder{x_i}&\pder{y_i}\end{bmatrix}^\mathrm{T}$; The two-particle interaction potential $U_2(\norm{\bm{x}_i-\bm{x}_j})$ governs particle interactions, with $\norm{\cdot}$ denoting the Euclidean norm; The spatiotemporal-varying external field induces an angular velocity $\Omega_i=\Omega(\bm{x}_i,\theta_i, t)$; $D_\mathrm{R}=3D_\mathrm{T}/a^2$ represents the rotational diffusion coefficient, where $a$ is the particle diameter; Thermal fluctuations are modeled through Wiener processes $\bm{W}_{\mathrm{T},j}(t)$ and $W_{\mathrm{R},j}(t)$ with zero-mean and unit variance white noise statistics.

For particle interactions, we employ the Weeks-Chandler-Andersen potential: $U_2(r) = \left(4\epsilon\left[\left(\frac{a}{r}\right)^{12}-\left(\frac{a}{r}\right)^6\right]+\epsilon\right)H(2^{1/6}a-r)$, where $\epsilon$ represents the interaction strength, $r$ denotes the particle separation distance, and $H(\cdot)$ is the Heaviside step function. The external field aligns particles in the $x$ direction according to $\Omega(\bm{x},\theta, t) = -\Omega_x(x,t) \sin(\theta)$, where positive values of $\Omega_x$ promote alignment in the $+x$ direction and vice versa. The field maintains uniformity in $y$ and is constrained to be periodic in $x$ with period $W$ and maximum magnitude $\Omega_{\max}$, such that $\norm{\Omega_x(\bm{x},t)}_\infty \leq \Omega_{\max}$.

We nondimensionalize \cref{eq:stoc_gov1,eq:stoc_gov2} using the rotational diffusion time $\taur=1/D_\mathrm{R}$, run length $\ell=U_0\taur$, and thermal energy $k_\mathrm{B}T$ as our fundamental units. This yields two key dimensionless parameters: the ratio of run length to microscopic diffusion length $(\ell/\delta)^2= U_0^2\taur/D_\mathrm{T}$, where $\delta=\sqrt{D_\mathrm{T}/\taur}$, and the area fraction $\phi= \bar{n}\pi a^2/4$, with $\bar{n}$ representing the spatially averaged particle number density.

For our simulations, we set $N=10^4$, $(\ell/\delta)^2=100$, $\phi=0.4$, $W=10\ell$, and $\Omega_{\max}=3\taur$. The simulation width is set to $W$ to match the external field's periodicity, and the height is adjusted to ensure the correct area fraction. Under these conditions, the system exhibits MIPS in the absence of an external field \cite{cates:tailleur:2015}. The dynamics are simulated using HOOMD-blue \cite{anderson:glaser:glotzer:2020}.
\subsection{Deriving vs. Learning Continuum-level Models}\label{sec:smol}
Rigorously, to obtain a continuum-level model, one could begin with the $N$-body Smoluchowski equation and account for binary interactions with an interaction potential that depends only on the particle separation distance:
\begin{gather}
	\dot{P}_N(\bm{X},t) +\sum_{i=1}^N \nabla_{i}\cdot \bm{j}_i + \pder{\theta_i}j_{\mathrm{R},i} = 0\\
	\bm{j}_i = U_0\bm{q}_iP_N(\bm{X},t)-D_\mathrm{T}\nabla_{i} P_N(\bm{X},t)+\nu_{\mathrm{int},i} P_N(\bm{X},t)\\
	\bm{j}_{\mathrm{R},i} = \Omega_i P_N(\bm{X},t) - D_\mathrm{R}\pder{\theta_i}P_N(\bm{X},t)
\end{gather}
$P_N(\bm{X},t)$ is the $N$-particle distribution function; $\bm{X} = \begin{bmatrix}\bm{x}_1^T,\ldots,\bm{x}_N^T, \theta_1,\ldots,\theta_N\end{bmatrix}^T$ denotes the collection of the particles' positions and orientations; $\bm{j}_i$ and $\bm{j}_{\mathrm{R},i}$ are the translational flux and rotational flux associated with particle $i$, respectively. $\nu_{\mathrm{int},i}$ represents the inter-particle interaction term; and $\Omega_i$ is the angular velocity induced by the external field on particle $i$.

The traditional approach of integrating over $N-1$ particle positions and orientations yields the Bogoliubov-Born-Green-Kirkwood-Yvon (BBGKY) hierarchy, where the one-particle distribution function $P_1(\bm{x},\theta,t)$ depends on the two-particle distribution function $P_2(\bm{x}_1,\theta_1,\bm{x}_2,\theta_2,t)$ and so on. The primary challenge lies in determining a closure relation that links higher-order distribution functions to lower-order ones \cite{hansen:mcdonald:2013,vrugt:bickmann:wittkowski:2023,saintillan:shelley:2013}. One closure strategy is to rewrite the two-particle distribution function in terms of the pair-distribution function. Assuming weakly varying external fields, the pair-distribution function can be learned using data from microscopic simulations \cite{jeggle:stenhammar:wittkowski:2020,bickmann:broker:tevrugt:wittkowski:2023}. This reduces the hierarchy to focus on the one-particle distribution function.

To make further progress with the one-particle distribution function, it is often expanded as a series of orientational moments, such as density, polar order, nematic order, and higher-order terms \cite{saintillan:shelley:2013}. This expansion typically requires an additional moment closure to truncate the series. By assuming that the polar order dominates over higher-order moments and that it is quasi-stationary with respect to density changes, one arrives at a density-only active Model B+ \cite{vrugt:bickmann:wittkowski:2023}.

In our work, we aim to achieve closure at the density level only while accounting for external fields that may vary significantly in both space and time. 
As a proof of concept, we focus on the number density, i.e. the zeroth orientational moment. The dynamics of this quantity can be described by an advection-diffusion equation:
\begin{equation}
	\dot{n}(\bm{x},t) + \nabla\cdot\left(-D_\mathrm{T} \nabla n(\bm{x},t)+\bm{v}(\bm{x},t)n(\bm{x},t)\right) = 0
	\label{eq:advection_diffusion}
\end{equation}
$\bm{v}(\bm{x},t)$ represents the effective velocity field that encompasses self-propulsion, multibody particle interactions, and the external field. 

By writing \cref{eq:advection_diffusion}, we are ``closing'' the moments expansion at the zeroth moment (the density), but capturing all higher-order moments via the advective term, $\bm{v}(\bm{x},t)$.
This is the key concept that differentiates the UDE approach from prior analytical moments expansion approaches.
Similar to learning the pair-distribution function, our goal is to use data from particle-based simulations and later, experiments, to learn $\bm{v}(\bm{x},t)$ without directly dealing with the interactions terms and higher-order couplings analytically.
At the end of the day, we end up with a single advection-diffusion equation that can be used in various applications, including optimal control. 
While an exact mapping---relating measured number densities, external fields, and their past 
values to the effective velocity field---would enable MPC, such a mapping is not readily available.
To these ends, we employ a UDE to learn $\bm{v}(\bm{x},t)$. This method draws parallels to stochastic force inference or learning interaction potentials from stochastic trajectories, since we learn an effective force from stochastic simulations that captures the combined effects of interparticle interactions, external fields, and self-propulsion \cite{frishman:ronceray:2020,king:engel:schoenholz:manoharan:brenner:2025}. The resulting inferred model provides both predictive accuracy and a foundation for implementing MPC to control emergent behaviors.
\subsection{Forced Delay UDEs} \label{sec:gray}
Following \cite{lovelett:avalos:kevrekidis:2019,kumar:rawlings:2023a}, we consider a forced dynamical system of the form
\begin{align}
	\dot{X} &= F(X(t),\Omega(t))
\end{align}
$F$ describes dynamics of the full state $X$ and the external input $\Omega$. Now suppose we cannot directly measure $X(t)$ but rather some function of the full state $n(t) = \Phi(X(t))$, $\Phi(\cdot)$ is the measurement function, and we would like to develop a dynamical model for the measurable $n(t)$. Furthermore, we have knowledge of some physics that constrain how $n(t)$ can evolve, e.g. conservation laws or symmetry, expressed as $f(\cdot)$.
\begin{equation}
	\dot{n}(t) = f(n(t),\Omega(t), v(X(t), \Omega(t)))
	\label{eq:pre-embedding}
\end{equation}
$v(X(t), \Omega(t))$ is the unknown function of the full state $X(t)$ that evolves the dynamics of $n$ exactly. However, we can only measure $n(t)$ which may only be a partial measurement of the full state. Motivated by Taken's embedding theorem, we introduce a delay embedding $p(t) = \begin{bmatrix}n(t-h), \Omega(t-h), \ldots, n(t-(N_p-1)h), \Omega(t-(N_p-1)h)\end{bmatrix}^\mathrm{T}$ where $h$ is the sampling time and $N_p$ is the embedding dimension \cite{takens:1981}. We assume that there exists an estimator $\Psi$ that estimates the full state $X(t)$ from the measurement $n(t)$ and its past $p(t)$.
For practical usage, $N_p$ is heuristically chosen to provide a reasonable fit to the data. 
Thus, we can rewrite \cref{eq:pre-embedding} as
\begin{equation}
	\dot{n}(t) = f(n(t),\Omega(t), \hat{v}(n(t), \Omega(t), p(t)))
	\label{eq:post-embedding}
\end{equation}
where the last unknown is the function $\hat{v}(\cdot)$ that handles both the estimation of the full state with estimator $\Psi$ and evaluates the effect of the full state on the state of interest $n(t)$. Thus, we approximate $\hat{v}(\cdot)$ with a parameterized function approximator, namely a feedforward artificial neural network $v_\mathrm{NN}(n(t), \Omega(t), p(t);\beta_\mathrm{NN})$ where $\beta_\mathrm{NN}$ are the network parameters.
\begin{equation}
	\dot{n}(t) = f(n(t),\Omega(t), v_\mathrm{NN}(n(t), \Omega(t), p(t);\beta_\mathrm{NN}))
	\label{eq:gray-box}
\end{equation}

This UDE modeling framework naturally extends to partial differential equations through a combination of spectral methods and the method of lines \cite{patel:trask:wood:cyr:2021,rackauckas:ma:skinner:ramadhan:edelman:2021}. Specifically, spectral methods allow us to represent the spatial components of the PDE using basis functions with time-varying coefficients. The method of lines then transforms the PDE into a system of ODEs governing these spectral coefficients. In this setting, the neural network becomes a neural operator that maps the current and past spectral coefficients (through the delay embedding) to their future evolution.

\subsection{Training data}\label{sec:train_d}
  With measurements of $n(t)$ available at sampling time $h$ and zero-order hold applied to $\Omega_x$, we can represent this as a discrete-time dynamical system with time step $k$. 
For simplicity, we refer to the measurement at time $k$ as $n(k)$ and the external field at time $k$ as $\Omega_x(k)$.
The model is designed to accurately represent both transient dynamics and steady-state behavior. The external field is subject to constraints $\norm{\Omega_x(\bm{x},t)}_\infty \leq \Omega_{\max}$. To systematically explore the external field's capabilities up to the maximum magnitude $\Omega_{\max}$, we employ spatial collocation at uniform intervals in $x$. At these collocation points, we sample the external field values uniformly from the interval $[-\Omega_{\max}, \Omega_{\max}]$.
At the initial time step and at the end of each hold period, we iterate the following:
\begin{enumerate}
\item Sample new external fields at collocation points uniformly from $[-\Omega_{\max}, \Omega_{\max}]$, and a new hold duration $k_h$ uniformly from integers in $[k_{\min}, k_{\max}]$.
\item Hold the external field constant at $\Omega_x$ for the next $k_h$ time steps, or until total time steps $k_f+N_p$, while recording both the number density $n(k)$ and external field $\Omega_x(k)$ at each step.
\end{enumerate}
At the end of data collection, we have a trajectory of measurements and inputs which we split into initial conditions for the past $p_0=(n(0),\Omega_x(0)\ldots,n(N_p-1),\Omega_x(N_p-1))$ and the training trajectory number densities $\underline{n} = (n(N_p), \ldots,n(N_p+k_f))$ and inputs $\underline{\Omega} = (\Omega_x(N_p), \ldots, \Omega_x(k_f+N_p-1))$. 

We repeat this process for $N_\mathrm{traj}$ trajectories to obtain a dataset of initial conditions and training trajectories.
 \subsection{Training protocol}\label{sec:train_p}
 For clarity, we introduce the multi-step prediction error minimization approach with a single trajectory.
 To obtain the unknown parameters $\beta_\mathrm{NN}$, we minimize the squared norm between the measured and predicted trajectories.
 \begin{mini*}|s|
	{\beta_\mathrm{NN}}{\sum_{k=0}^{k_f}\norm{\underline{n}(k)-\hat{n}(k)}^2}
	{}{}
	\addConstraint{\hat{n}(k+1) = }{f_h(\hat{n}(k),\underline{\Omega}(k),v_\mathrm{NN}(\hat{n}(k), \underline{\Omega}(k), p(k);\beta_\mathrm{NN}))}
	\addConstraint{\hat{p}(k+1) = }{f_p(\hat{n}(k),\underline{\Omega}(k),\hat{p}(k))}
	\addConstraint{\hat{n}(0) =}{\underline{n}(0)}
	\addConstraint{\hat{p}(0) =}{p_0}
\end{mini*}
where $f_h(\cdot)$ is the discrete time dynamics of the number density $n$ obtained via numerical integration of \cref{eq:gray-box} with step size $h$, e.g., a Runge Kutta method;
$f_p(\cdot)$ is a sliding window function that drops the oldest element of the past vector $\hat{p}(k)$ and appends the newest elements, namely, $\hat{n}(k)$ and $\underline{\Omega}(k)$;
In other words, provided initial conditions and the external field sequence, a predicted trajectory is generated by integrating the UDE \cref{eq:gray-box} with the estimated parameters $\beta_\mathrm{NN}$ and the error between the predicted and measured trajectories is minimized by updating the parameters $\beta_\mathrm{NN}$. This approach has been shown to increase the long time accuracy as compared to a single step approach, i.e. $\min_{\beta_\mathrm{NN}}\sum_{k=0}^{k_f-1}|\underline{n}(k+1)-f(\underline{n}(k),\underline{\Omega}(k),v_\mathrm{NN}(\underline{n}(k), \underline{\Omega}(k), \underline{p}(k);\beta_\mathrm{NN}))|^2$ \cite{maddu:weady:shelley:2024,askham:kutz:2018}.
For $N_\mathrm{traj}$ trajectories, we minimize the sum of the trajectory errors to find the optimal parameters $\beta_\mathrm{NN}$.

\subsection{Learning the advective flux}\label{sec:learn_flux}
We numerically approximate the advection-diffusion equation \cref{eq:gray-box-n} using spectral methods.
\begin{align*}
	\dot{\bm{n}} &= -\iota\bm{K}(-D_\mathrm{T}\iota\bm{K}+\bm{T}\bm{v}_\mathrm{NN}(\bm{n},\bm{\Omega},\bm{p};\beta_\mathrm{NN}))\bm{n}\\
	&\coloneqq \bm{F}(\bm{v}_\mathrm{NN}(\bm{n},\bm{\Omega},\bm{p};\beta_\mathrm{NN}))\bm{n}
\end{align*}
$\iota$ is the imaginary number; $\bm{n}\in\mathbb{R}^{d_n}$ is the vector of Fourier coefficients for the number density;
$\bm{\Omega}\in\mathbb{R}^{d_\Omega}$ is vector of external field coefficients;
$\bm{p}\in\mathbb{R}^{N_p(d_n+d_\Omega)}$ is the vector of past number density and external field coefficients;
$\bm{v}_\mathrm{NN}(\cdot):\mathbb{R}^{(N_p+1)(d_n+d_\Omega)}\rightarrow \mathbb{R}^{d_v}$ is the neural network that maps the number density, external field coefficients, and the past density and external field coefficients to the velocity field coefficients and is parameterized by $\beta_\mathrm{NN}$;
 $\bm{K}$ is the Fourier differentiation matrix;
 $\bm{T}$ is the rank 3 tensor that Toeplitzizes the neural network output $\bm{v}_\mathrm{NN}(\cdot)$ for convolution with $\bm{n}$.
 
 Provided $N_\mathrm{traj}$ trajectories of length $k_f$ with initial number density and past vectors, we minimize the sum of trajectory errors to find the optimal parameters $\beta_\mathrm{NN}$:
 \begin{mini!}|s|
	{\beta_\mathrm{NN}}{\sum_{i=1}^{N_\mathrm{traj}}\sum_{k=0}^{k_f}|\bm{\underline{n}}_i(k)
		-\hat{\bm{n}}_i(k)|^2}
		{}{}
	\label{eq:trajectories-a}
	\addConstraint{\hat{\bm{n}}_i(k+1) = }{\hat{\bm{n}}_i(k)+h\bm{F}(\hat{\bm{v}}(k))\hat{\bm{n}}_i(k+1)}
	\label{eq:trajectories-b}
	\addConstraint{\hat{\bm{v}}(k) = }{\bm{v}_\mathrm{NN}(\hat{\bm{n}}_i(k),\bm{\underline{\Omega}}_i(k),\hat{\bm{p}}_i(k);\beta_\mathrm{NN})}
	\label{eq:trajectories-c}
	\addConstraint{\hat{\bm{p}}_i(k+1) = }{(\hat{\bm{n}}_i(k),\bm{\underline{\Omega}}_i(k),\hat{\bm{p}}_i^{0:N_p-2}(k))}
	\label{eq:trajectories-d}
	\addConstraint{\hat{\bm{n}}_i(0) =}{\bm{\underline{n}}_i(0)}
	\label{eq:trajectories-e}
	\addConstraint{\hat{\bm{p}}_i(0) =}{\bm{p}_{i,0}}
	\label{eq:trajectories-f}
\end{mini!}
 $\bm{\underline{n}}_{i}(k)$ and $\hat{\bm{n}}_{i}(k)$ are the measured and predicted density trajectory for the $k$th step of the $i$th trajectory, respectively;
$\bm{\underline{\Omega}}_i(k)$ is the external field at the $k$th step in the $i$th trajectory;
$\hat{\bm{p}}_i(k)$ is the predicted past vector at step $k$ of the $i$th trajectory;
$\bm{p}_{i,0}$ is the initial past density and external field coefficients for the $i$th trajectory. 

\cref{eq:trajectories-a} is the sum of squared errors between the predicted and measured number density trajectories;
\cref{eq:trajectories-b} is the numerical integration of the advection-diffusion equation with a modified backward Euler method;
\cref{eq:trajectories-c} is the neural network approximation for the effective velocity;
\cref{eq:trajectories-d} describes the sliding window for the past coefficients;
$\bm{p}_i^{0:N_p-2}$ is the vector of past coefficients excluding the oldest set of coefficients;
\cref{eq:trajectories-e,eq:trajectories-f} are the initial conditions for the number density and past coefficients, respectively.

Note the diffusion operator is stiff, which requires either small time steps or an implicit solver. We would like large time steps for control, which motivates an implicit solver. Thus, the numerical integration of the advection-diffusion equation is done using a modified backward Euler method where the nonlinear term is evaluated using only the current time step. This allows \cref{eq:trajectories-b} to be linear in $\hat{\bm{n}}_i(k+1)$. This approach is analogous to an implicit-explicit method where nonlinear non-stiff portions are evaluated explicitly while linear stiff operations are handled implicitly. Rather than have a separate dynamical system for $\bm{v}_\mathrm{NN}$ which we evolve separately, we simply use the past to predict the future velocity field which minimizes errors in number density trajectories. 
The main motivation for the modified approach is to avoid tracking gradients through an iterative nonlinear solver. With the implicit-explicit approximation, solving \cref{eq:trajectories-b} requires a single Newton step as the problem is linear. Alternatively, adjoint sensitivities could be used, but we found the modified backward Euler method to be faster for our problem \cite{chen:rubanova:bettencourt:duvenaud:2018}. 
Alternatively, one could leverage the implicit function theorem to define the solver's dynamics in terms of its sensitivity to inputs and directly compute gradients for optimization \cite{look:doneva:kandemir:gemulla:peters:2020}.

We used $N_\mathrm{traj}=1000$, $k_f=20$, $N_p=2$, 
$d_n=d_v=21$, $d_\Omega = 11$, $h=0.5\taur$, $k_{\min}=5$ and $k_{\max}=20$.
For the neural network, we used a single-layer feedforward neural network with 64 nodes and hyperbolic tangent activation functions. Optimization was done using the Adam optimizer in PyTorch with a learning rate of $10^{-3}$.

In addition to the training data, we used a validation set of 100 trajectories to monitor the training process. 
More specifically, we used the validation set to monitor the loss function and if the validation loss did not decrease for 20 epochs, we stopped training.

Lastly, the testing data, i.e., data not used during any part of training, we chose to evaluate the model on $N_\mathrm{test}=100$ trajectories of length $k_f=200$.
We evaluated a longer trajectory to ensure the model had long time prediction capabilities.

\subsection{Orthogonal series estimation}\label{sec:orth}
We use an orthogonal series estimator to estimate the density.
Suppose you have a random variable $x$ with density $n(x)$ and you wish to estimate
$n(x)$ from $N$ samples $\begin{bmatrix}x_1 & x_2 & \cdots & x_N\end{bmatrix}^\mathrm{T}$.
We use the following estimator.
\begin{align*}
	\hat{n}(x) = \sum_{j=-d}^d \hat{a}_j P_j(x)\\
	\hat{a}_j = \frac{1}{N\norm{P_j}^2}\sum_{i=1}^N P_j(x_i)
\end{align*}
$\hat{n}(x)$ is the orthogonal series estimate;
$P_j(x)$ being the $j$th orthogonal basis function;
$\hat{a}_j$ being the $j$th coefficient.
$d$ is the degree of the orthogonal series.
In other words, the true coefficients are the expected value of the basis functions, i.e., $a_j = \langle P_j(x)\rangle$, 
so we estimate the coefficients by averaging over the samples.

This can also be used to estimate averages of observable quantities $A(x)$, e.g., fluxes or polar orders.

\begin{align*}
	\hat{A}(x) = \sum_{j=-d}^d \hat{a}_{A,j} P_j(x)\\
	\hat{a}_{A,j} = \frac{1}{N\norm{P_j}^2}\sum_{i=1}^N P_j(x_i)A_i
\end{align*}
$A_i$ is the observable quantity for sample $i$.
For the examples in the main paper, we use a Fourier basis, i.e., $P_j(x) = e^{2\pi \iota j x/W}$.
To handle confinement in examples in the SI, we use Legendre polynomials.
\subsection{Stage cost formulation} 
\subsubsection*{Notation}
We introduce some notation for the stage costs. The operator $\langle \cdot \rangle$ denotes the expectation of its argument over the entire domain, evaluated using number density $n(x,t)$, i.e.,
\begin{equation*}
	\langle \cdot \rangle = \int (\cdot) n(x, t) \, \mathrm{d}x
\end{equation*}
To represent the probability of finding a particle in the right half of the domain, we define 
$n(x>0)= \langle H(x)\rangle$. Similarly, the probability of finding a particle in the left half of the domain is denoted as $n(x<0)$.
The operator $\langle \cdot \rangle_{\mathcal{D}}$ denotes the expectation of its argument over subdomain $\mathcal{D}$, also evaluated using number density $n(x,t)$. For example, the expectation restricted to the right half of the domain is expressed as:
\begin{equation*}
	\langle \cdot \rangle_{x>0} = \frac{\langle(\cdot)H(x)\rangle}{n(x>0)}
\end{equation*}
\subsubsection{Stage cost for splitting and juggling population}\label{sec:stg_cost1}
Recall the control targets are to accumulate particles about $x_+$ and $x_-$ and to match target left-to-right density ratios. To achieve this, we associate a cost with each objective and define the stage cost $L_1(\cdot)$ as follows:
\begin{gather}
	L_1(n(x,t),r_{\mathrm{sp}}(t)) =  c_1\left(L_+ + L_- \right)
	+ c_2 L_r
	\label{eq:dance_cost}
	\\
	L_- = \Bigl<(x-x_-)^2\Bigr>_{x<0}\\
	L_+ = \Bigl<(x-x_+)^2\Bigr>_{x>0}\\
	L_r = \left(n(x<0)- r_\mathrm{sp}(t)n(x>0)\right)^2
\end{gather}
In the stage cost $L_1(\cdot)$, we penalize only number density deviations, as opposed to also penalizing a control effort. 
The cost $L_1$ consists of three contributions. 
$L_-$ penalizes the expected square distance of particles left of the origin to the left set point $x_-$.
Likewise, $L_+$ penalizes the expected square distance of particles right of the origin to the right set point $x_+$.
The last term in the cost $L_r$ penalizes deviations from the left-to-right ratio set point $r_\mathrm{sp}(t)$ 
between probability of finding a particle to the left and right of the origin.
This ratio penalty is used to achieve set points like equal-sized groups or 30\% of particles to the left. 

The weights for the positional set point penalty and the ratio penalty are chosen to be $c_1=1$ and $c_2=100$, respectively.
$c_2$ was chosen large since the scale of the ratio penalties are much smaller compared to the positional set point penalties.
The terminal cost $L_f$ is chosen to be the same as $L_{1}$. 
\subsubsection{Stage cost for simultaneous density and flux control}\label{sec:stg_cost2}
To simultaneously maintain the particles at $x_+$ and $x_-$, while prescribing the mean flux, the stage cost extends \cref{eq:dance_cost} by adding a term that penalizes deviations between measured and set point mean fluxes.
\begin{gather}
	L_2(n(x,t), \psi_\mathrm{sp}(t)) = L_1(n(x,t))+L_\psi \\
	L_\psi =c_3 \left(\psi-\psi_\mathrm{sp}(t)\right)^2,\quad \psi = \frac{\langle v_x \rangle}{W}\\
	\psi_\mathrm{sp}(t) = \psi_{\max}\sin\left(\frac{2\pi}{T}t\right)
\end{gather}
$\psi$ represents the mean flux (i.e., flux averaged over the domain); $c_3$ is the weight for the flux penalty; $\psi_{\max}$ is the target flux magnitude, and $T$ is the period. The terminal cost $L_f$ is chosen to be $L_{1}$. Since the percentage of particles in the left or right half is not relevant in this example, we set the ratio penalty $c_2=0$.

We set $\psi_{\max}=0.01\ell/\taur$, $T=20\taur$, $c_1=1$, and $c_3=10^3$. Starting with particles equally split into two populations, we simulate the system for $20\taur$.
\clearpage 
\twocolumngrid

\typeout{}
\bibliography{abbreviations,books,articles,unpub,proceedings,resgrpsubmitted}

\end{document}